\documentstyle[epsf,11pt]{article}

\newcommand{\Nf}{N_{\!f}} 
\newcommand{\kslash}{k \! \! \! /} 
\newcommand{\Dslash}{D \! \! \! \! /}

\newcommand{\MSbar}{\overline{\mbox{MS}}} 

\begin{document}

\title{QCD and QED renormalization group functions - a large $N_{\!f}$ 
       approach\thanks{Invited talk presented at DESY Workshop on QCD and QED 
                       in Higher Order, Rheinsberg, Germany, 21st-26th April,
                       1996.}
       } 

\author{J.A. Gracey\thanks{PPARC Advanced Fellow.}  
        \\ Department of Mathematical Sciences, University of
        Liverpool, \\ P.O. Box 147, Liverpool, L69 3BX, United Kingdom}  

\date{} 

\maketitle

\vspace{0.5cm} 
\noindent 
The large $N_{\!f}$ self-consistency programme is reviewed. As an application 
the QCD $\beta$-function is computed at $O(1/N_{\!f})$ and the anomalous 
dimensions of polarized twist-$2$ singlet operators are determined at the same 
order. 

\vspace{-8cm} 
\hspace{13.4cm} 
{\bf {LTH-369}} 
\vspace{8cm} 

\section{INTRODUCTION}

The principle of computing higher order perturbative corrections to quantities 
in a quantum field theory is simple to state. Specifically one wishes to 
determine the coefficients of a series expansion in powers of the coupling 
constant, $g$, which is assumed to be small. The difficulty is that each 
coefficient arises from a set of Feynman diagrams. Since the number of graphs 
increases rapidly with each order in $g$, this becomes a huge computational 
exercise. Indeed few calculations are available beyond the third order for four
dimensional gauge theories, [1-3]. Such computations, however, can be 
complemented by alternative methods. For example, if a quantity whose 
perturbative structure is required, depends on parameters other than the 
coupling constant, then it may be possible to perform some sort of expansion in
those other variables. In other words, to approach the calculation of the 
coefficients from another angle. Such a possibility exists for models where the
basic fields lie in a multiplet of an internal symmetry group like 
$SU(\Nf)$. Then an expansion in powers of the dimensionless variable $1/\Nf$ 
can be performed when $\Nf$ is large. Included in this class of theories are 
QED and QCD.   

\section{BASIC FORMALISM}

We can express this more concretely for such theories by considering the 
structure of a typical quantity, $F(g)$, whose expansion is of the form, 
\begin{equation} 
F(g) ~=~ a_1g ~+~ (a_2\Nf + b_1)g^2 ~+~ (a_3\Nf^2 + b_2\Nf + c_1)g^3 ~+~ O(g^4) 
\end{equation}  
The aim is to determine $\{a_i\}$, $\{b_i\}$ and $\{c_i\}$ and so on, with 
respect to some renormalization scheme. For this article we will consider only
the $\MSbar$ scheme. We have included the dependence on $\Nf$ explicitly. 
Perturbative calculations systematically extract $\{a_1\}$, $\{a_2,b_1\}$ and
$\{a_3,b_2,c_1\}$ and so on. However, $F(g)$ can equally be calculated as an
expansion in powers of $1/\Nf$ with $\Nf$ large provided $g\Nf$ is held fixed.
Such an approach is in effect a reordering of the set of Feynman diagrams 
making up $F(g)$. Clearly the coefficients which are computed successively at 
each order in $1/\Nf$ are $\{a_i\}$, $\{b_i\}$ and $\{c_i\}$ and so on. The 
question that now arises is how to determine an infinite set of coefficients in
a compact way.  

For the first part of this article we review the method to compute the 
structure of the functions of the renormalization group equation, (RGE). This 
is achieved by using ideas from statistical physics and the critical 
renormalization group and which were initially applied to the $O(N)$ $\sigma$
model in \cite{4}. Basically at a phase transition universal quantities can be 
measured experimentally or calculated theoretically which characterise the 
physics. These critical exponents are fundamental and are determined from the 
renormalization group functions evaluated at the fixed point. They are 
functions of the space-time dimension, $d$ $=$ $2\mu$, and the parameters of 
any internal symmetries of the underlying field theory, \cite{5}. For example, 
the exponent $\eta$ $=$ $\eta(d,\Nf)$ is related to the wave function 
renormalization. With this relation to the RGE functions, if one computes 
critical exponents directly, in say the $1/\Nf$ approximation, and provided one
knows the location of the critical coupling, $g_c$, as a function of $d$ in 
powers of $1/\Nf$, then one can deduce the coefficients of the RGE functions 
from an $\epsilon$-expansion of the exponent where $d$ $=$ $4$ $-$ $2\epsilon$. 

For QED and QCD this procedure is possible provided one considers each theory
in $d$-dimensions, when $\beta(g)$ has a non-trivial zero. Specifically, 
\cite{1},  
\begin{equation} 
\beta(g) = (d-4)g + \left[ \frac{2}{3} T(R)\Nf - \frac{11}{6} C_2(G) 
\right] g^2 
\end{equation} 
has a fixed point at, \cite{6},  
\begin{equation} 
g_c ~=~ 3\epsilon/[T(R)\Nf] ~+~ O(1/\Nf^2) 
\end{equation} 
where the colour group Casimirs are defined by $\mbox{tr} (T^a T^b)$ $=$ $T(R)
\delta^{ab}$, $T^a T^a$ $=$ $C_2(R)$ and $f^{acd}f^{bcd}$ $=$ $C_2(G)
\delta^{ab}$ with $T^a_{IJ}$ the generator of the colour group and $f^{abc}$ 
its structure constants. So if $F(g)$ is the quark wave function 
renormalization then 
\[ 
F(g_c) ~=~ \frac{1}{\Nf} \left[ \sum_{n=1}^\infty a_n (3\epsilon/T(R))^n 
\right] ~+~ O(1/\Nf^2) 
\] 
and the coefficients $\{a_i\}$ are clearly related to those of 
$\eta_1(\epsilon)$ $=$ $F(g_c)$ where $\eta$ $=$ $\sum_{n=1}^\infty 
\eta_n(\epsilon)/\Nf^n$.

Exponents such as $\eta$ are defined by considering the action as a 
dimensionless quantity at $g_c$. With the usual QCD lagrangian  
\begin{equation} 
L ~=~ i \bar{\psi}^{iI} \Dslash \psi^{iI} ~-~ (G^a_{\mu \nu})^2/4e^2 
\end{equation} 
where $1$ $\leq$ $i$ $\leq$ $\Nf$, $D_\mu$ $=$ $\partial_\mu$ $+$ $iT^a 
A^a_\mu$ and $G^a_{\mu\nu}$ $=$ $\partial_\mu A^a_\nu$ $-$ $\partial_\nu 
A^a_\mu$ $+$ $f^{abc}A^b_\mu A^c_\nu$, then the quark and gluon propagators at 
criticality will have the asymptotic form, \cite{6}, in an arbitrary covariant 
gauge with parameter $b$, as $k^2$ $\rightarrow$ $\infty$,  
\begin{eqnarray}
\psi(k) & \sim & \frac{A\kslash}{(k^2)^{\mu-\alpha}} \nonumber \\  
A_{\mu\nu}(k) & \sim & \frac{B}{(k^2)^{\mu-\beta}} \left[ \eta_{\mu\nu} 
{}~-~ (1-b)\frac{k_\mu k_\nu}{k^2} \right]  
\end{eqnarray} 
Here $A$ and $B$ are $k$-independent amplitudes and the anomalous parts of the
dimensions $\alpha$ and $\beta$ are defined relative to the canonical pieces as
$\alpha$ $=$ $\mu$ $-$ $1$ $+$ $\eta/2$ and $\beta$ $=$ $1$ $-$ $\eta$ $-$ 
$\chi$. The exponent $\chi$ is the quark gluon vertex anomalous dimension. 
Indeed each operator built out of the fields of (4) will have an associated 
critical exponent. 

Several leading order exponents have been determined in QCD. For example, in
the Landau gauge at leading order in $1/\Nf$, \cite{6},  
\begin{equation}
\eta ~=~ \frac{C_2(R)\eta^{\mbox{o}}_1}{T(R)\Nf} ~~~,~~~  
\eta ~+~ \chi ~=~ - ~ \frac{C_2(G) \eta^{\mbox{o}}_1}{2(\mu-2)T(R)\Nf} 
\end{equation}
where $\eta^{\mbox{o}}_1$ $=$ $(2\mu-1)(\mu-2)\Gamma(2\mu)/[4\Gamma^2(\mu) 
\Gamma(\mu+1) \Gamma(2-\mu)]$. These have been deduced respectively by 
substituting the critical propagators (5) into the quark and gluon $2$-point
functions and the quark gluon $3$-point function and examining the scaling 
behaviour of the resulting integrals in the critical region. The expansion of
(6) in powers of $\epsilon$ agrees exactly with the explicit $3$-loop 
perturbative calculation of the corresponding renormalization group functions
in the Landau gauge, \cite{3}. One feature which leads to a simplification in 
these calculations is that one does not need to consider any diagrams where 
there are radiative corrections on an internal line. The reason is that one is 
using propagators which have non-zero anomalous dimensions. As these 
represent the effect of such corrections, including them in a calculation 
would in effect be a double counting. For higher order $1/\Nf$ corrections 
this reduces the number of diagrams that need to be computed at criticality.  

\section{QCD $\beta$-FUNCTION}

We now consider the computation of the QCD $\beta$-function as an application 
of the above formalism. First, we need to recall another simplification of the 
critical point approach. Sometimes more than one field theory underlies the 
description of a phase transition. In this case there is a choice of models 
which can be used to determine the critical exponents and such models are said 
to be in the same universality class. An example of this is the equivalence of 
the $O(N)$ $\sigma$ model and $\phi^4$ theory with an $O(N)$ symmetry. In three 
dimensions the exponents calculated in either determine the critical behaviour 
in the Heisenberg ferromagnet. However, if one model in the universality class 
has a simpler structure then an exponent calculation could be substantially 
reduced by computing with it. This is the case for (4). In the large $\Nf$ 
limit it has been shown, \cite{7}, that QCD is equivalent to the non-abelian 
Thirring model, (NATM), which has the lagrangian,   
\begin{equation} 
L ~=~ i \bar{\psi}^{iI} \Dslash \psi^{iI} \, - \, (A^a_\mu)^2/2\lambda 
\end{equation} 
where $\lambda$ is the coupling constant which is dimensionless in two 
dimensions. It is more efficient to use (7) to deduce universal critical 
exponents due to the absence of the additional triple and quartic gluon 
self-interactions. It was also shown in \cite{7}, however, that these 
interactions are correctly recovered in integrating quarks out of the gluon 
$3$- and $4$-point functions. Of course, (3) must be used to determine the 
perturbative coefficients of quantities in QCD, \cite{6}. Also, as we are 
considering a non-abelian gauge theory in covariant gauges the ghost sector 
must be added to both  lagrangians. At leading order in $1/\Nf$, though, we do 
not need to consider contributions from graphs with ghosts as they are 
supressed by an additional factor of $1/\Nf$. 

To find $\beta(g)$ at $O(1/\Nf)$ we use a scaling law between various 
anomalous dimensions to determine the non-zero exponent $\omega$ $=$ $-$ 
$\beta^\prime(g_c)/2$. It is deduced from the gluon kinetic term of (4), as
\begin{equation} 
\omega ~=~ \eta ~+~ \chi ~+~ \chi_G 
\end{equation} 
where the first two terms arise from the gluon fields in the field strength 
composite operator $G$ $=$ $(G^a_{\mu\nu})^2$ which has anomalous dimension 
$\chi_G$. It is determined by examining a Green's function with $G$ as an 
insertion and, using (5), calculating the leading order $1/\Nf$ set of 
contributing graphs. These are illustrated in figure 1. 
\begin{figure} 
\vspace{0.5cm} 
\epsfxsize=12cm 
\epsfbox{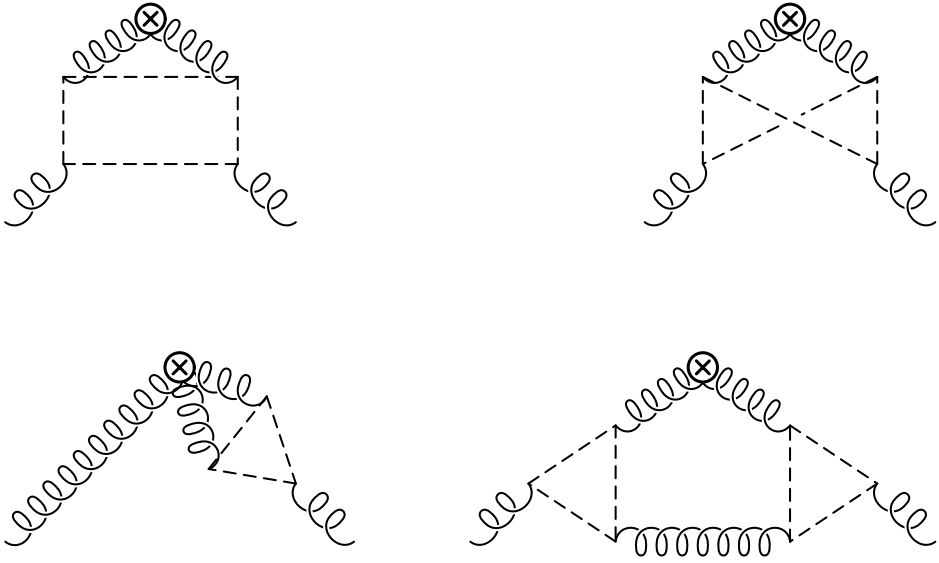} 
\caption{Graphs for $\omega_1$.} 
\end{figure}
In \cite{9}, however, the structure of the QED $\beta$-function at $O(1/\Nf)$ 
was determined by explicitly carrying out the leading order bubble sum and 
using $\MSbar$ renormalization. This was later extended to the critical 
formalism in \cite{10}, which corresponds to the calculation of the first two 
graphs of figure 1. Therefore for the non-abelian extension one need only 
compute the remaining two graphs which involve the colour group factor 
$C_2(G)$. Although the critical exponent associated with the $\beta$-function 
is independent of the gauge parameter, to avoid mixing with the operator 
$(\partial^\mu A^a_\mu)^2$, \cite{11}, we compute $\omega$ in the Landau gauge. 
With $b$ $=$ $0$ we find, \cite{8},  
\begin{equation} 
\omega_1 ~=~ -~ \frac{C_2(G)\eta^{\mbox{o}}_1}{T(R)} \left[ (2\mu-3)(\mu-3) 
\frac{C_2(R)}{C_2(G)} ~-~ \frac{(4\mu^4 - 18\mu^3 + 44\mu^2 - 45\mu + 14)}
{4(2\mu-1)(\mu-1)} \right] 
\end{equation} 
where $\omega_0$ $=$ $(\mu-2)$. The $\epsilon$-expansion of (9) correctly 
reproduces the $3$-loop (gauge independent) result of [1-3] and allows us to 
deduce new coefficients at higher orders. In the notation of (1),  
\begin{eqnarray} 
a_4 &=& - ~ [154C_2(R) ~+~ 53C_2(G)]/3888 \\ 
a_5 &=& C_2(R)(144\zeta(3) ~+~ 107)/15552 
{}~+~ C_2(G)(480\zeta(3) ~-~ 229)/31104  
\end{eqnarray}  
where $\zeta(n)$ is the Riemann $\zeta$-function. 

We close this section by noting some technical points in the determination of 
$\chi_G$ and anomalous dimensions of other operators. The contribution from 
each graph is found from the residue of the simple pole in the regulator of the
critical theory, \cite{12}. This (analytic) regularization is introduced by 
shifting the dimension of the gluon field by an infinitesimal amount, $\Delta$.
In other words we calculate the graphs using (5) but with the replacement 
$\beta$ $\rightarrow$ $\beta$ $-$ $\Delta$. It is important to appreciate that 
the dimension of space-time is fixed in these exponent calculations, unlike in 
explicit perturbative computations using dimensional regularization. 

\section{POLARIZED OPERATORS}

We have extended this analysis to other operators whose renormalization is 
necessary for the operator product expansion used in deep inelastic scattering.
For instance, one can study the twist-$2$ flavour singlet polarized operators, 
\cite{13}, 
\begin{eqnarray} 
{\cal O}_G &=& \frac{i^n}{2} {\cal S} \epsilon^{\mu_1\alpha\beta\gamma} 
\mbox{tr} G_{\beta\gamma} D^{\mu_2} \ldots D^{\mu_{n-1}} 
G^{\mu_n}_{~~\, \alpha} \nonumber \\  
{\cal O}_F &=& i^n {\cal S} \bar{\psi} \gamma^5 \gamma^{\mu_1} D^{\mu_2} \ldots
D^{\mu_n} \psi  
\end{eqnarray} 
where ${\cal S}$ denotes symmetrization in the Lorentz indices and subtraction 
of all combinations of traces and $n$ is the moment of the operator. The 
treatment of unpolarized flavour non-singlet and singlet operators at 
$O(1/\Nf)$ was given in \cite{14}. We recall that the gauge independent 
anomalous dimension of these physical operators is responsible for the way in
which the associated Wilson coefficients vary with energy. It is important to 
have large $\Nf$ information available on the renormalization of (12) as the 
matrix of anomalous dimensions, $\gamma_{ij}(g)$, has recently been evaluated 
at second order, \cite{15,16}. A matrix of renormalization constants arises in 
perturbation theory due to the mixing under renormalization of both operators 
as they have the same quantum numbers and canonical dimensions, \cite{14}.  

The computation of the corresponding exponent matrix, $\gamma_{ij}(g_c)$, is by
the technique outlined in the previous section. We substitute the operators in
Green's functions and determine the scaling behaviour. In the large $\Nf$ 
approach there is a simplification. At criticality, due to the field content, 
the operators have different canonical dimensions and therefore do not mix. 
Alternatively this can be seen by calculating $\gamma_{ij}(g_c)$ directly and 
observing that it is triangular at $O(1/\Nf)$, \cite{14}. 

In order to relate exponent results to perturbation theory one observes that 
the exponent derived from the insertion of ${\cal O}_F$ actually corresponds to
the eigenoperator of the perturbative mixing matrix which is predominantly 
fermionic in its content. In other words working at $g_c$ one circumvents the 
need to construct operators which diagonalize the RGE governing the evolution 
of the fermionic and gluonic singlet Wilson coefficients. The result of our 
calculation is, for even $n$,  
\begin{eqnarray} 
\eta^{(n)}_{{\mbox{\footnotesize{F}}},1} &=& 
\frac{2C_2(R)\eta^{\mbox{o}}_1}{(2\mu-1)(\mu-2)T(R)} \left[ \frac{}{} (\mu-1)^2 
{}~-~ \frac{\mu(\mu-1)^3}{(\mu+n-1)(\mu+n-2)} \right. \nonumber \\
&&\left. +~ 2\mu(\mu-1) [\psi(\mu-1+n) - \psi(\mu)]  
{}~-~ \frac{\mu(2\mu+n-5)(n+2)S(n)}{2(\mu+n-1)(\mu+n-2)} \right] 
\end{eqnarray} 
where $S(n)$ $=$ $\Gamma(n)\Gamma(2\mu)/\Gamma(2\mu+n-2)$ and $\psi(x)$ is the
logarithmic derivative of the $\Gamma$-function. It agrees with the two loop
results of \cite{15,16} when $\gamma_{ij}(g)$ is diagonalized. The leading 
order $1/\Nf$ value of the dimension of ${\cal O}_G$ is equivalent to the one 
loop value, which follows from the $\Nf$ structure of the entries in 
$\gamma_{ij}(g)$. For the sake of comparison we note the exponent for the 
corresponding unpolarized operator ${\cal O}_f$ $=$ $i^n {\cal S} \bar{\psi} 
\gamma^{\mu_1} D^{\mu_2}$ $\ldots$ $D^{\mu_n} \psi$ is, \cite{14},  
\begin{eqnarray}
\eta^{(n)}_{{\mbox{\footnotesize{f}}},1} &=& 
\frac{(\mu-1)C_2(R)\eta^{\mbox{o}}_1}{(2\mu-1)
(\mu-2)T(R)\Nf} \left[ \frac{}{} 2(\mu-1) 
{}~ -~ \frac{2\mu(\mu-1)^2}{(\mu+n-1)(\mu+n-2)} \right. \nonumber \\
&&+~ \left. 4\mu[\psi(\mu-1+n) - \psi(\mu)] \right. \nonumber \\
&&-~ \left. \frac{\mu(2\mu-1+n)\Gamma(n-1)\Gamma(2\mu)} 
{(\mu+n-1)(\mu+n-2) \Gamma(2\mu+n)} \right. \nonumber \\ 
&&~~~~ \times \left. [(n(n-1)+2(\mu-1+n))^2  
+ 2n(n-1)(\mu-2)(2\mu-3+2n) \right. \nonumber \\ 
&&~~~~~~ \left. +~ 4(\mu-2)(\mu-1+n)] \frac{}{} \right]
\end{eqnarray}

The purely four dimensional object $\gamma^5$ needs to be carefully treated in 
the large $\Nf$ method, \cite{17}. Unlike in some approaches in dimensional 
regularization we can use a fully anticommuting $\gamma^5$ in $d$-dimensions 
since our regularization is achieved by a shift in the gluon dimension, 
\cite{18}. For closed fermion loops one performs the integral in $d$-dimensions
before projecting with, for example, \cite{17},  
\begin{equation} 
\mbox{tr}(\gamma^5 \gamma^\mu \gamma^\nu \gamma^\sigma \gamma^\rho) ~=~ 
4 \epsilon_{\mu\nu\sigma\rho} 
\end{equation}  
 
As a final application we have also computed the anomalous dimension of the 
singlet axial current, ${\cal O}_5$ $=$ $\bar{\psi} \gamma^\mu \gamma^5 \psi$. 
The renormalization of this operator is more involved as it is not conserved in
the quantum theory due to the chiral anomaly, \cite{19}. However, in 
perturbation theory methods have been developed to obtain the three loop 
anomalous dimension in the $\MSbar$ scheme. (See, for example, \cite{20}.) 
Briefly this is a two part exercise. One part involves carrying out a standard 
$d$-dimensional renormalization using dimensional regularization. A finite 
renormalization is also required. This is determined by ensuring the one loop 
character of the operator form of the anomaly is preserved, \cite{20}. A 
similar two part approach in $1/\Nf$, using the above rules for graphs 
involving $\gamma^5$, yields the critical dimension of ${\cal O}_5$ as, 
\cite{18},   
\begin{equation}
\eta_{5,1} ~=~ - ~ \frac{3\mu C_2(R) \eta^{\mbox{o}}_1}{(\mu-1)T(R)} 
\end{equation} 
The $\epsilon$-expansion of (16) agrees with the three loop result of 
\cite{21,20}. 

\section{STRUCTURE FUNCTIONS} 

A second area of application of $1/\Nf$ methods is in the determination of the
higher order structure of the finite parts of Green's functions or amplitudes.
Not only are such calculations important in obtaining new coefficients of a 
series, they can also be used to study summability problems. As the $1/\Nf$ 
approach is a reordering of perturbation theory in which graphs with infinite
chains of quark loops are treated first, one could substitute such chains into 
the appropriate amplitude and carry out the renormalization to leave a finite
result. In practice this will be tedious and inefficient. For a better approach
we recall a feature of perturbative calculations. In dimensionally regularized
calculations with massless fields, the effect of performing a quark loop 
integral is to obtain a $d$-dependent result with a factor which involves the 
momentum raised to a $d$-dependent power. In the renormalization process this 
latter part is important in obtaining the finite piece. Motivated by this and 
the simple structure of the fixed point propagators, we can obtain $O(1/\Nf)$ 
information on the finite part efficiently by using the propagators, \cite{14}, 
\begin{eqnarray}
\psi(k) &=& \frac{\kslash}{k^2} \nonumber \\  
A_{\mu\nu}(k) &=& \frac{1}{(k^2)^{1+\delta}} \left[ \eta_{\mu\nu} 
{}~-~ \frac{k_\mu k_\nu}{k^2} \right]  
\end{eqnarray} 
We stress that this approach is not at a fixed point and is closer in spirit
to perturbative calculations. The gluon propagator has an adjusted power, 
$\delta$, where $\delta$ $=$ $4\Nf T(R)g/3$. Calculations are performed by
substituting (13) into Green's functions with undressed gluon lines with the
integration restricted to four dimensions. The resulting $\delta$-dependent
function can be expanded in powers of $\delta$ and it transpires that the 
coefficients are in direct correspondence with those obtained in perturbation
theory. 

Several applications are available. For instance, the finite part of the photon
propagator has been determined in QED, \cite{22}. In deep inelastic scattering 
the $O(1/\Nf)$ part of the longitudinal non-singlet structure function has been
studied. The Wilson coefficients at the $L$th loop is, \cite{14},   
\begin{equation}  
C^{\mbox{\footnotesize{NS}}}_{\mbox{\footnotesize{long}}}(1,g,L) ~=~ 
\frac{d^L~}{d\delta^L} \! 
\left[ \frac{8C_2(R) e^{5\delta/3}\bar{S}(n,\delta)g}{(2-\delta)(1-\delta) x^n}
\right]  
\end{equation}
where $\bar{S}(n,\delta)$ $=$ $\Gamma(n+\delta)/[\Gamma(n)\Gamma(1+\delta) 
/(n+1+\delta)]$, and $x$ is the Bjorken scaling variable. It agrees with the
three loop results of \cite{23}. 

The poles in $\delta$ at $\delta$ $=$ $1$ and $2$ correspond to renormalons in 
the series. Recently, these have been used with the naive non-abelianization
procedure, \cite{24}, to estimate the effect of higher twist contributions to
the full structure function, \cite{25}  

\section{CONCLUSIONS} 

We have focussed primarily on the $O(1/\Nf)$ corrections to the RGE functions.
One important feature of the fixed point approach, however, is that 
$O(1/\Nf^2)$ calculations are also possible in $d$-dimensions. For example, the 
anomalous dimension of the electron mass in QED has been determined at this
order, \cite{26}. With the simplification that occurs from the equivalence of 
QCD and NATM the extension of this and other results will be possible for the 
non-abelian case.

\end{document}